# Observation of single antiferromagnetic magnon modes through tunnelling spectroscopy of spin-1/2 Kitaev system $\alpha$-$RuCl_3$

Servet Ozdemir,[1, a] Mikhail Kashchenko,[1, b] and Kostya S. Novoselov[1, c]
*Department of Physics and Astronomy, University of Manchester, Manchester, UK*

(*Electronic mail: s.ozdemir@leeds.ac.uk)

The small-gap room-temperature semiconductor $\alpha$-$RuCl_3$, which is known to undergo a Mott-Hubbard transition at low temperatures, is one of the most promising candidates for realisation of an exotic matter form, the quantum spin liquid state, which may have applications in quantum computing. Although extensively investigated by neutron scattering techniques, electronic study of this system in the form of van der Waals heterostructures has been limited mainly to graphene proximity. Here we report a systematic study of planar and tunnelling electronic properties of $\alpha$-$RuCl_3$ films, where we observe an *n*-type field effect on $\alpha$-$RuCl_3$ films at room temperature, with a Mott insulator nature onset below 120 K. For films of three-layer thickness and below we find inelastic scattering features, below the Néel temperature of 7-14.5 K, which we attribute to single magnon modes. Our study confirms preserved low-temperature signatures of the zigzag antiferromagnetic order in the atomically thin limit and its single magnon modes within the continuum through tunnelling spectroscopy.

In the early studies since the 1960s[1], layered van der Waals $\alpha$-$RuCl_3$ crystals were established as narrow-gap room-temperature semiconductors, with *n*-type conduction properties at room temperature where activation energies range from 0.1 to 0.3 eV.[2,3] In the following years, band gaps measured in spectroscopy experiments seemingly influenced by both charge and spin interactions led to an understanding of the system as a Mott-Hubbard insulator.[4] A more contemporary set of transport experiments utilising Hall bars fabricated on mechanically exfoliated $\alpha$-$RuCl_3$ films suggested a phase transition around 180 K, with the crystals ceasing to have any conductive properties below 120 K, hence electrically confirming the Mott-insulating nature of $\alpha$-$RuCl_3$ and its temperature onset.[5] Although the symmetry of the room temperature bulk crystal structure is being debated,[6–9] a structural phase transition in $\alpha$-$RuCl_3$ crystals have been reported by multiple groups,[7,10,11] and the Mott nature of the insulator transition is due to the spin orbit coupling effects emerging in the lower temperature phase.[12] Furthermore, a low-temperature magnetic phase transition, which is accepted to be of zigzag antiferromagnetic order, has been observed with the Néel temperature being reported at 7-14 K,[7,10,11] way below the reported structural transitions above 100 K.

A surge in interest in $\alpha$-$RuCl_3$ was triggered when, in 2006, Kitaev showed that in spin ½ honeycomb systems, topological properties applicable to quantum computation emerge, in the currently called quantum spin liquid state.[13] The emergence of resonant valence bond states was hypothesised in understanding spin-orbit coupling-induced Mott transition in $\alpha$-$RuCl_3$; hence $\alpha$-$RuCl_3$ became one of the most promising candidates for a realised spin ½ Kitaev system.[12,14] Most promising physical signatures of the model Hamiltonian of the system, the competing zigzag antiferromagnetic order emerging at lower temperatures, and the Majorana excitations of the quantum spin liquid state at intermediate temperatures were attributed to the experimental results obtained in neutron scattering studies.[6,15,16] Similar signatures were also measured by thermal conductivity experiments, accompanied by a thermal quantum Hall effect.[17–19] Breakthroughs in van der Waals magnetic insulators have enabled studies of magnetic phenomena in tunnelling devices of van der Waals heterostructures.[20–22] Despite $\alpha$-$RuCl_3$ being a layered van der Waals material, experimental tunnelling spectroscopy device investigation into potential exotic physics displayed by exfoliated $\alpha$-$RuCl_3$ crystals has been very limited.[23,24] Existing tunnelling studies are mostly scanning tunnelling microscopy studies[25–27]. Majority of the device work on $\alpha$-$RuCl_3$ has been proximity studies of graphene or other crystals to majorly reasonably thick ($> 10\,\mathrm{nm}$) $\alpha$-$RuCl_3$ flakes with accepted consensus being that a large charge transfer takes place, in case of graphene, leaving it *p*-doped[28–32].

In this paper, we report both planar transport and tunnelling experiments on exfoliated $\alpha$-$RuCl_3$ crystals. $\alpha$-$RuCl_3$ flakes were encapsulated in hBN crystals, and graphene electrodes were utilised for the tunnelling devices. At room temperature, exfoliated $\alpha$-$RuCl_3$ crystals are conductive; they were measured with a 4-probe geometry and exhibit an *n*-type field effect consistent with *n*-type transport reported in earlier literature.[3] We find that below 120 K the Mott insulator transition is indeed manifested with absence of any possible planar transport.[5] We show through tunnelling spectroscopy that the insulating nature of $\alpha$-$RuCl_3$ increases as the temperature drops below 120 K, with tunnel junctions becoming more resistive at lower temperatures. In contrast to previous reports on tunnelling devices that only showed a broadened excitation continuum above 4 meV,[23] we find reproducible single magnon mode features in 1 to 3-layer tunnel junctions above this energy range but within the continuum as theoretically

[a] Present address:School of Physics and Astronomy, University of Leeds, Leeds, UK
[b] Present address:Programmable Functional Materials Lab, Center for Neurophysics and Neuromorphic Technologies, 127495 Moscow, and Moscow Center for Advanced Studies, Kulakova str. 20, 123592 Moscow, Russia
[c] Present address:Institute for Functional Intelligent Materials, National University of Singapore, 117544 Singapore, Singapore

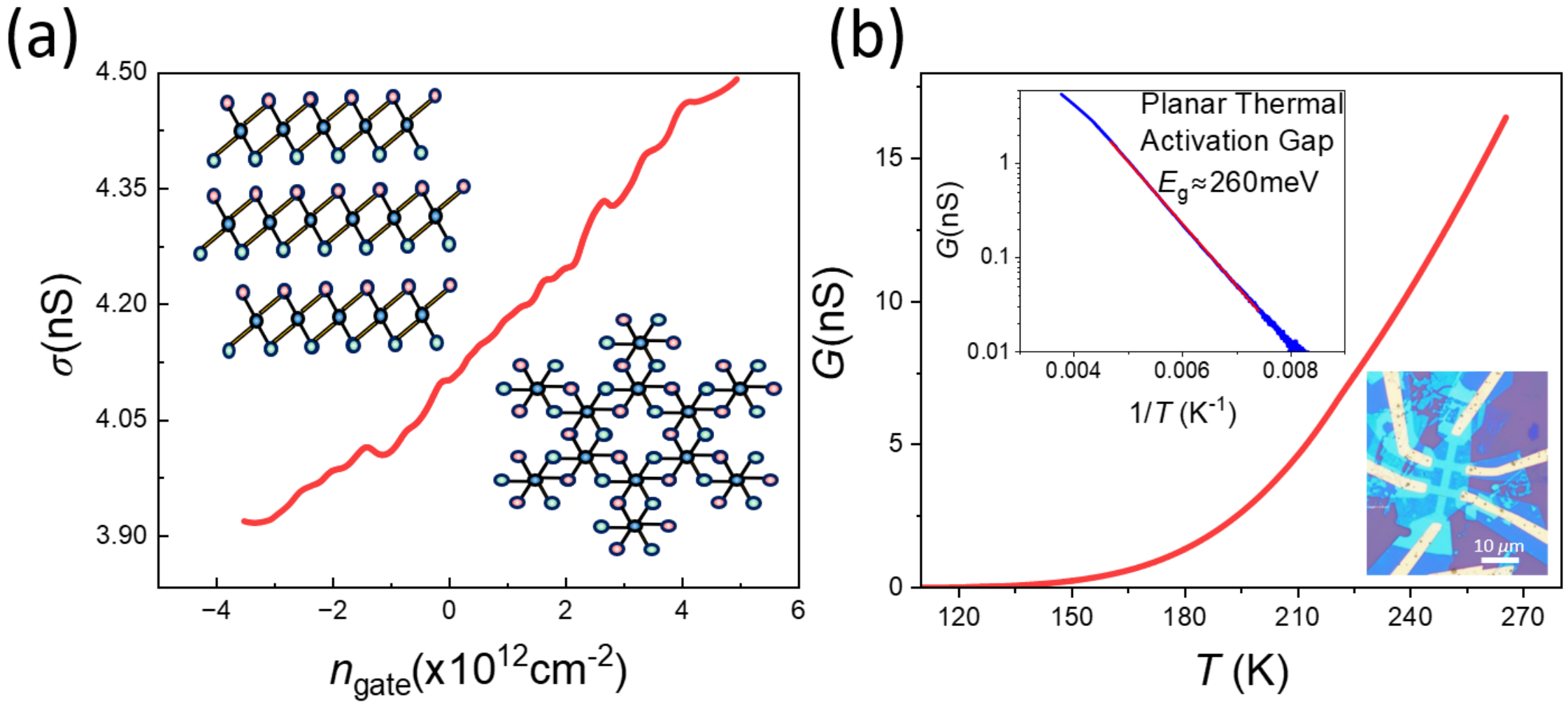

FIG. 1. (a) Room temperature conductivity measured as a function of backgate induced carrier density, indicating an electric field effect on a 4-terminal tri-layer planar contacted $\alpha$-$RuCl_3$ film suggesting n-type conduction yielding a field effect mobility of $\approx 1\times10^{-3}\,cm^2/Vs$ with layered crystal structure of $\alpha$-$RuCl_3$ depicted in upper inset and honeycomb structure on lower inset. (b) Planar cooling curve obtained on a tri-layer device with conduction totally absent at 120 K, suggesting a strong insulating nature below this temperature, with an Arrhenius plot shown in the inset yielding a thermal activation gap of $\approx 260\,meV$.

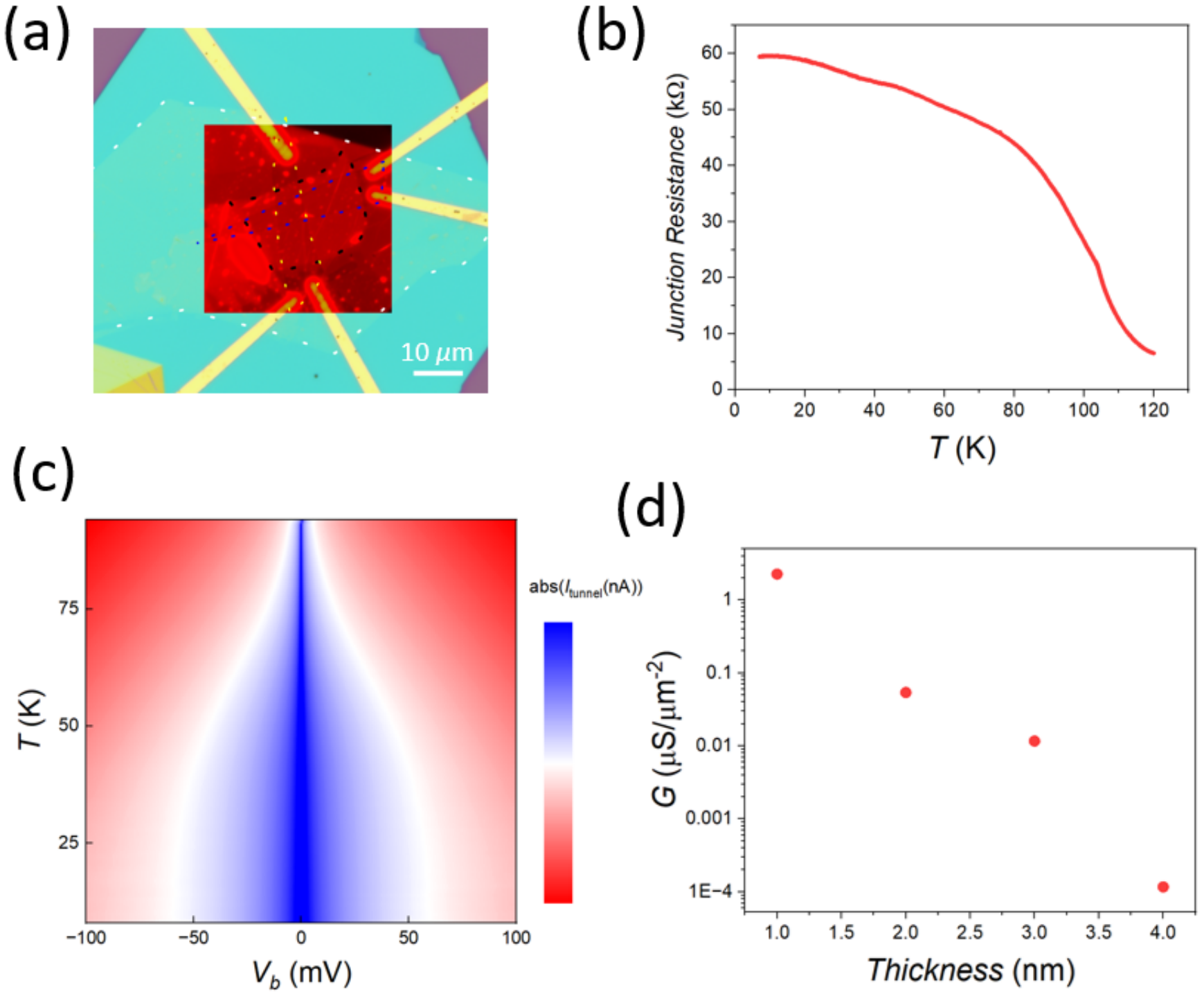

FIG. 2. (a) Optical micrograph of a bilayer tunnel junction device with top and bottom graphene electrodes. (b) Temperature dependence of tunnel junction resistance suggesting onset of the tunnelling behaviour at 120 K. (c) A map of temperature dependence of tunnelling $I$-$V$ curves suggesting an increasingly resistive Mott insulator (scale blue – red, 1 nA – 2 $\mu$A ). (d) Layer number dependence of low-temperature zero-bias $\alpha$-$RuCl_3$ tunnel junction resistance.

expected.[33]

Before device fabrication, both the bulk $\alpha$-$RuCl_3$ crystals (see Fig. 1a insets for crystal schematics) and the exfoliated flakes in ambient conditions (see Supplementary Material, Fig. S1) were characterised by Raman spectroscopy[34] (see Supplementary Material Fig. S2). Similarly to $\alpha$-$RuCl_3$ crystals, graphene electrodes in tunnelling devices were identified by optical contrast, confirmed by Raman spectroscopy and as-sembled into a stack (see Methods in Supplementary Material for details). Initially, planar devices of $\alpha$-$RuCl_3$ crystals were fabricated and characterised at room temperature. A weak yet $n$-type field effect was observed on $\alpha$-$RuCl_3$, in agreement with Rojas and Spinolo[3], who have previously observed electron-dominated transport in $\alpha$-$RuCl_3$. The field effect mobility was extracted to be very small, $\approx 1 \times 10^{-3}\,\mathrm{cm}^2/\mathrm{Vs}$, with the conductivity data being shown in Fig. 1a (see Supplementary Material, Fig. S3 for conductivity plotted as a function of gate voltage as well as gating set-up schematic). Further to this, temperature dependence of conductance was studied on various devices, and a total absence of conductance was observed below 120 K as shown in Fig. 1b. Fitting an Arrhenius thermal activation model (see Supplementary Material) to the temperature dependence of conductance, Fig. 1d inset, yields an activation gap of $\approx 260\,\mathrm{meV}$, again in agreement with the range of values reported in transport and optical characterisation of bulk films.[2,3]

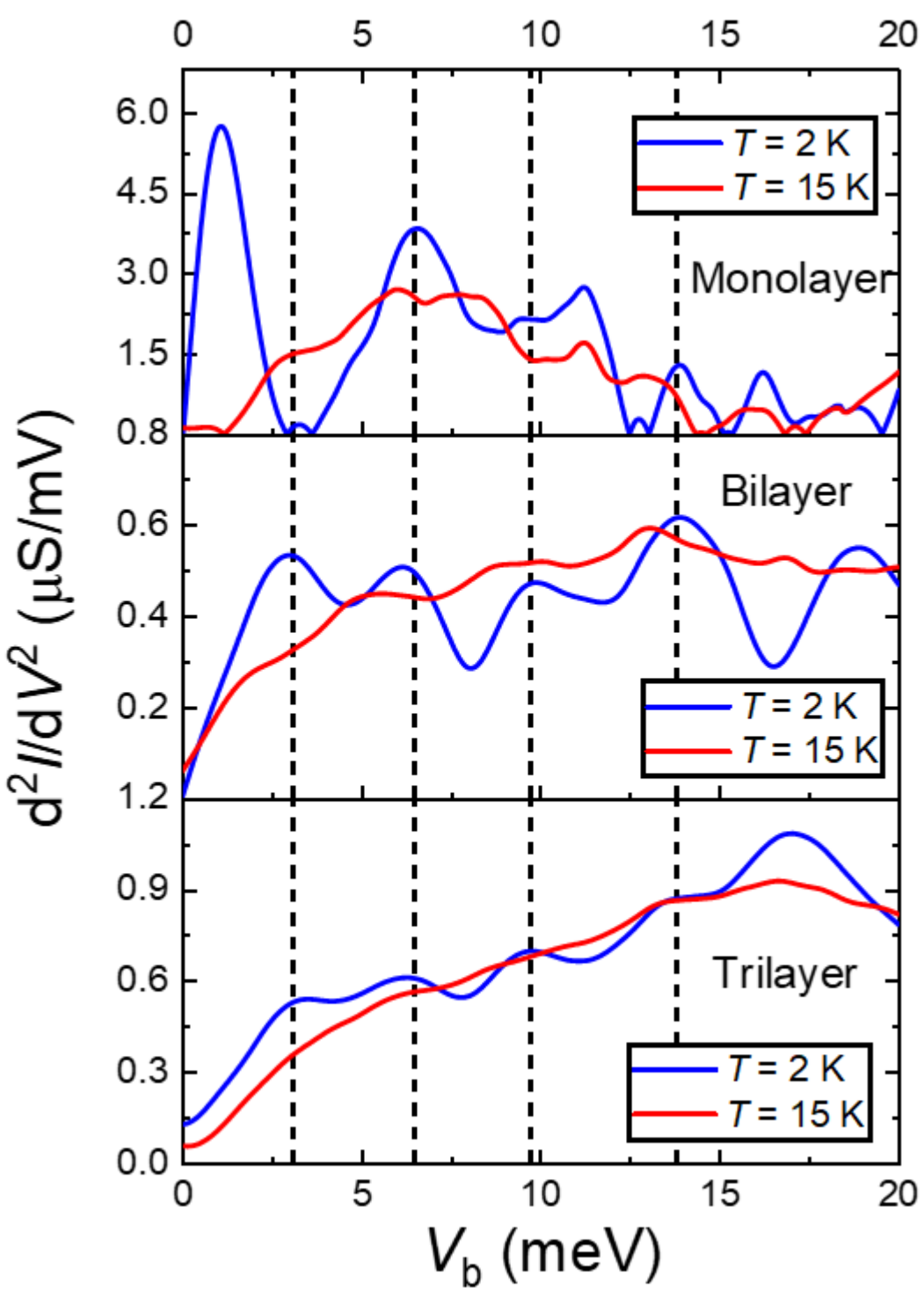

FIG. 3. Mono- (top), bi- (middle) and tri- (bottom) layer tunnel junction second derivative *I*-*V* curves below (blue) and above Néel temperature (red), suggesting the presence of single magnon mode related features below the Néel temperature, with 4 features common to all samples highlighted with dashed lines.

With the use of planar contacts being limited to higher temperatures, to gain more insights into magneto-electronic properties at lower temperatures, exfoliated $\alpha$-$RuCl_3$ were assembled into tunnelling devices[35] sandwiched between graphene flakes as shown in the micrograph on Fig. 2a. The tunnelling onset on these devices was found to take place at around 120 K (Fig. 2b), coinciding with the temperature at which the planar conductivity of $\alpha$-$RuCl_3$ flakes ceases. A map of the *I*-*V* curves is obtained as temperature is lowered, suggesting an enhancement in the apparent insulating nature of $\alpha$-$RuCl_3$ due to increasing tunnel junction resistance, which tends to be saturating below 25 K as shown in Fig. 2c. Low-temperature junction conductances are shown as a function of $\alpha$-$RuCl_3$ tunnel barrier thickness, where large differences in zero-bias conductance normalised per unit area are visible as shown in Fig. 2d, as is the case for other exfoliated van der Waals magnets[20,21]. We find that for $\alpha$-$RuCl_3$ films of thickness of 4 layers and above, tunnelling current is largely screened at low bias, hence prohibiting the study of low-energy magnon modes (see Massicotte et al.,[24] and Supplementary Material, Fig. S4).

As one of the prominent signatures of the below $7-14\,\mathrm{K}$ zigzag antiferromagnetic order, bulk spectroscopy experiments have revealed low-energy magnons within a broader excitation continuum.[6,15,16,36] Inelastic scattering events, including magnon-assisted ones, emerge in tunnelling spectroscopy experiments as features in the second derivative of the tunnel current *I*-*V* curves.[20,37] To investigate the expected emergence of the zigzag antiferromagnetic phase through tunnelling spectroscopy, we have studied temperature-dependent evolution of the second derivative of tunnelling *I*-*V* curves above and below the Néel temperature on mono-, bi- and trilayer films as shown in Fig. 3. Strong inelastic scattering features are present at $T = 2\,\mathrm{K}$ on all three films, with the continuum clearly being visible for bi- and tri-layer junctions. What is very clear is that sharp features observed at $T = 2\,\mathrm{K}$ on top of the continuum either disappear or merge into the continuum at $T = 15\,\mathrm{K}$, with common individual modes present across all three thicknesses of films at $T = 2\,\mathrm{K}$ depicted through dashed lines. For the monolayer tunnel barrier, the observation of such an early-onset additional peak at 1.03 mV as well as a differing appearance of the continuum compared to bilayer and trilayer tunnel barriers could be attributed to the buckling effects reported in the earlier literature[8,23]. We find that the observed individual modes on the monolayer device are highly reproducible, but the broad shape of the continuum is found to differ from device to device (see Supplementary Material, Fig. S5 for a comparison of two monolayer tunnelling devices).

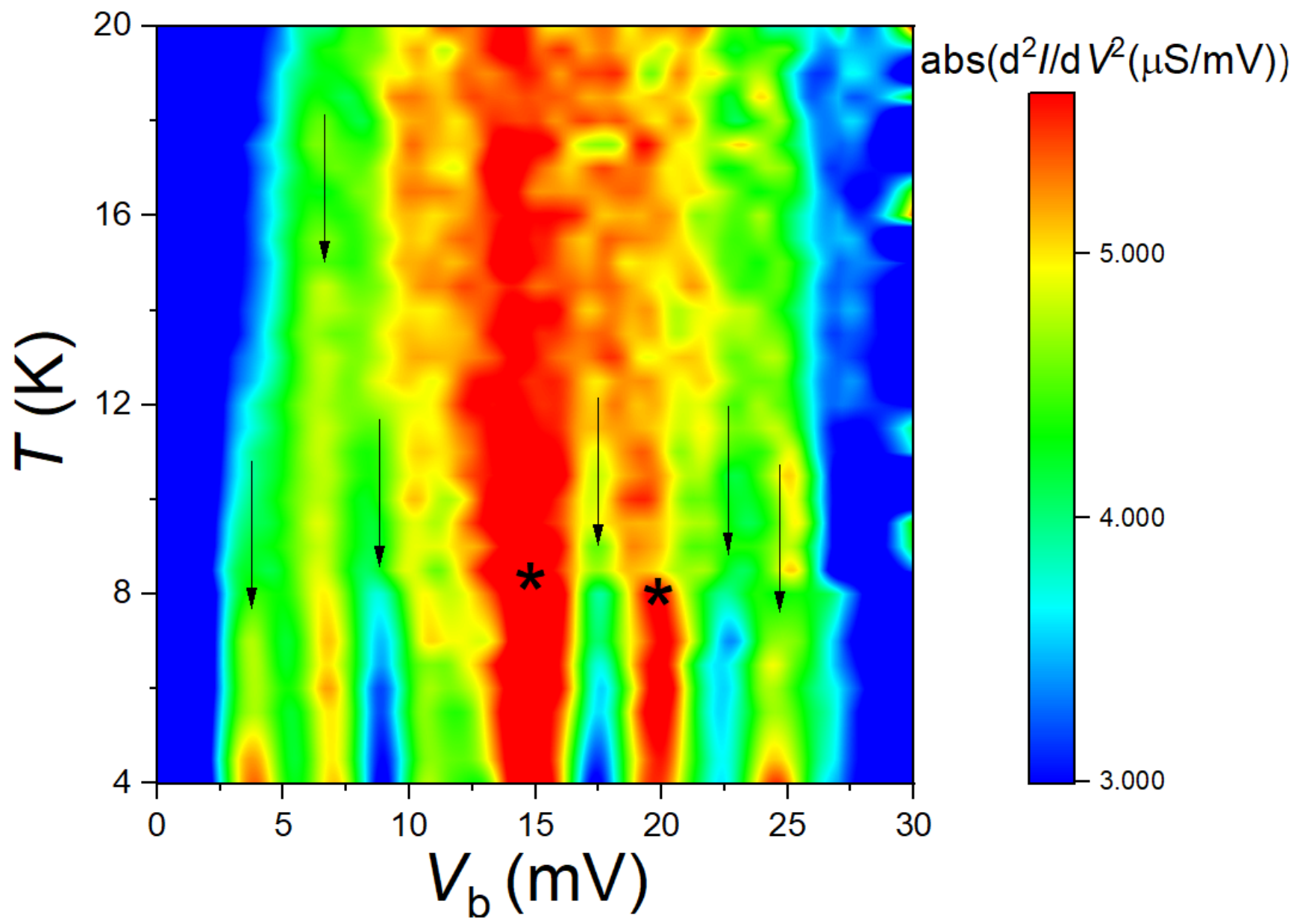

FIG. 4. Temperature dependence map of the second derivative of tunnel current on a bi-layer tunnel junction showing disappearance of single magnon mode features with temperature (pointed by the arrows) and prominence of phonon-related features (marked with * symbols) present until maximum measured temperature, 20K.

To be able to elaborate more on the nature of inelastic scattering features, we present a map of the second derivative of the tunnelling curves as a function of temperature, as shown in Fig. 4 (see Supplementary Material, Fig. S6 for the monolayer tunnelling device maps). A non-universal point about the zigzag-AFM order in $\alpha$-$RuCl_3$ has been the observed Néel temperature, which has been reported to be in the range of $7-14\,\mathrm{K}$, with the enhancement in Néel temperature towards the upper limit usually being attributed to the stacking faults.[12,38,39] Inelastic scattering events indicated by arrows disappear in the vicinity of the Néel temperature range, hence could be attributed to single magnon mode assisted tunnelling despite having a weak backgate dependence (see Supplementary Material, Fig. S7)[21]. The features marked with stars, around 14 mV and 20 mV, which develop into a broader single feature at higher temperatures, have been resolved in Raman spectroscopy characterisation of the exfoliated flakes (see Supplementary Material, Fig. S2 as well as literature[5,40]), and hence can be attributed to phonon-assisted tunnelling processes. Despite the striking difference between phonon-assisted tunnelling events and the magnon features marked with arrows, given their magnetic origin, the question of the magnetic field evolution of features marked with arrows arises. The magnetic field dependence of the derivative of the differential conductance for the same bilayer device is shown in Supporting Material, Fig. S8. As expected, phonon-assisted features have no dependence on the applied magnetic field. The magnon-assisted features have a complex field dependence, in agreement with existing studies where a non-linear dispersion as a function of magnetic field[41], disappearance[36], appearance at a finite magnetic field[36,42], as well as crossover[43] have been found. An additional point is that the low-energy features that are dependent on both temperature and magnetic field are found to persist until an energy value of 25 mV, in contrast to the commonly reported value of 15 meV for magnons in bulk systems.[44]

Overall, we have studied electronic properties of exfoliated mono-, bi- and tri-layer $\alpha$-$RuCl_3$ films in both planar and tunnelling contact architectures. We have found that at room temperature $\alpha$-$RuCl_3$ flakes display an $n$-type electric field effect by electrostatic gating, in agreement with $n$-type semiconductor observations in bulk.[3] We find that planar conductive property of $\alpha$-$RuCl_3$ thin films vanishes below temperature of $120\,\mathrm{K}$. This is accompanied by a temperature-dependent tunnel junction resistance onset of the films at $120\,\mathrm{K}$, with an onset of the Mott insulating behaviour of the films below this temperature. We have reported consistent inelastic scattering features within the continuum in the second derivative $I$-$V$ curves below a Néel temperature range of 7 to $14.5\,\mathrm{K}$. Through temperature dependence maps (as well as magnetic field), we distinguish the features that are likely to be due to single-magnon mode-assisted tunnelling and phonon-assisted tunnelling. Our study shows that exfoliated atomically thin films of $\alpha$-RuCl$_3$ still possess the experimental signatures of the zigzag antiferromagnetic order in the form of single-magnon modes within the continuum, where, in proximity to it, the quantum spin liquid state and Majorana excitations are expected to exist.

## SUPPLEMENTARY MATERIAL

Supplementary Material containing methods, characterisation of exfoliated flakes, and further data tunnelling spectroscopy data, including magnetic field dependence of inelastic scattering features is available online.

## ACKNOWLEDGMENTS

Servet Ozdemir acknowledges a PhD studentship from the EPSRC-funded Graphene NOWNANO CDT. Mikhail Kashchenko acknowledges the support of his work by RSF No. 25-29-00636.

## DATA AVAILABILITY STATEMENT

Data are available from the corresponding author upon reasonable request.

# Supplementary Material for Observation of single antiferromagnetic magnon modes through tunnelling spectroscopy of spin-1/2 Kitaev system α-$RuCl_3$

## Methods

**Fabrication**

Flakes of a-$RuCl_3$ crystals were exfoliated onto PPC coated $SiO_2$/Si substrates in ambient conditions. Varied thicknesses of a-$RuCl_3$ flakes were identified through optical contrast. For tunnelling devices, graphene electrodes were also identified through optical contrast and confirmed by Raman spectroscopy. Both stacks of hBN/a-RuCl3/hBN and hBN/Graphene/a-RuCl3/Graphene/hBN were assembled by the dry transfer method utilising a PMMA membrane. Cr/Au contacts were made for both planar and graphene tunnel electrode devices after a combination of $CHF_3$ and $O_2$ etching on a Reactive Ion Etcher system.

**Experimental characterisation**

Measurements were carried out using standard lock-in techniques with DC sourcemeters for AC-DC differential conductance measurements. Temperature dependence studies were carried out utilising continuous flow liquid helium systems.

**Gate-induced carrier density – capacitive gating formula**

The back gate-induced carrier density per unit area in electric field effect measurement was calculated using the formula

$$n_{gate} = \frac{V_{bg}\varepsilon_r\varepsilon_0}{ed}$$

where *d* is the thickness of the dielectric material, $\varepsilon_r$ is the relative permittivity of the dielectric material, $\varepsilon_0$ is the permittivity of free space, $V_{bg}$ is the applied gate voltage, and *e* is the charge of an electron.

**Arrhenius Law used to extract transport activation band gap**

To estimate the observed thermal activation gap, the Arrhenius equation

$$G = Aexp\left(\frac{-E_g}{2k_BT}\right)$$

was used, where G is conductance, A is a pre-exponential factor, $k_B$ is the Boltzmann constant, and $E_g$ is the activation gap, which is extracted from the data.

# Supplementary Figures

a

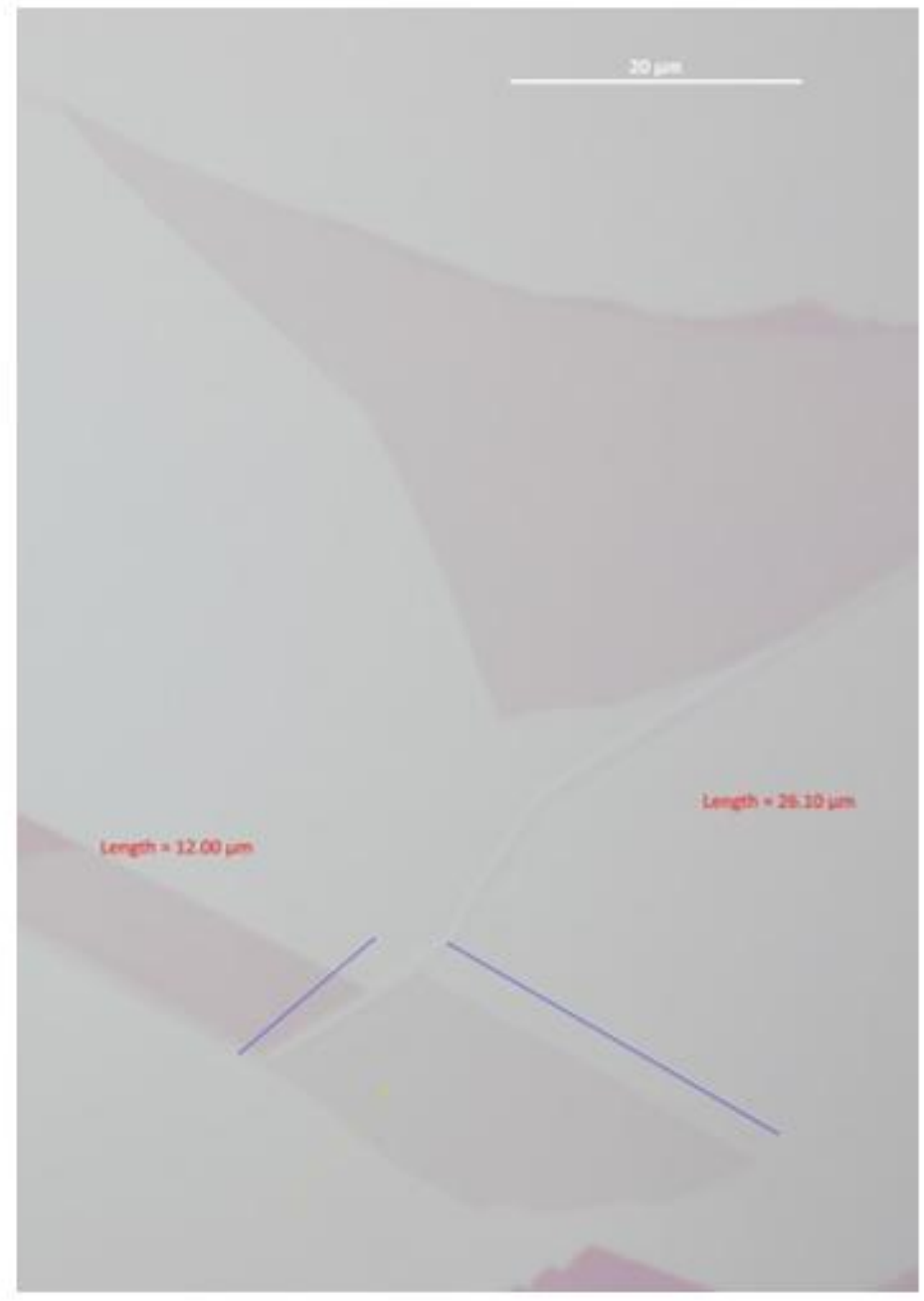

b

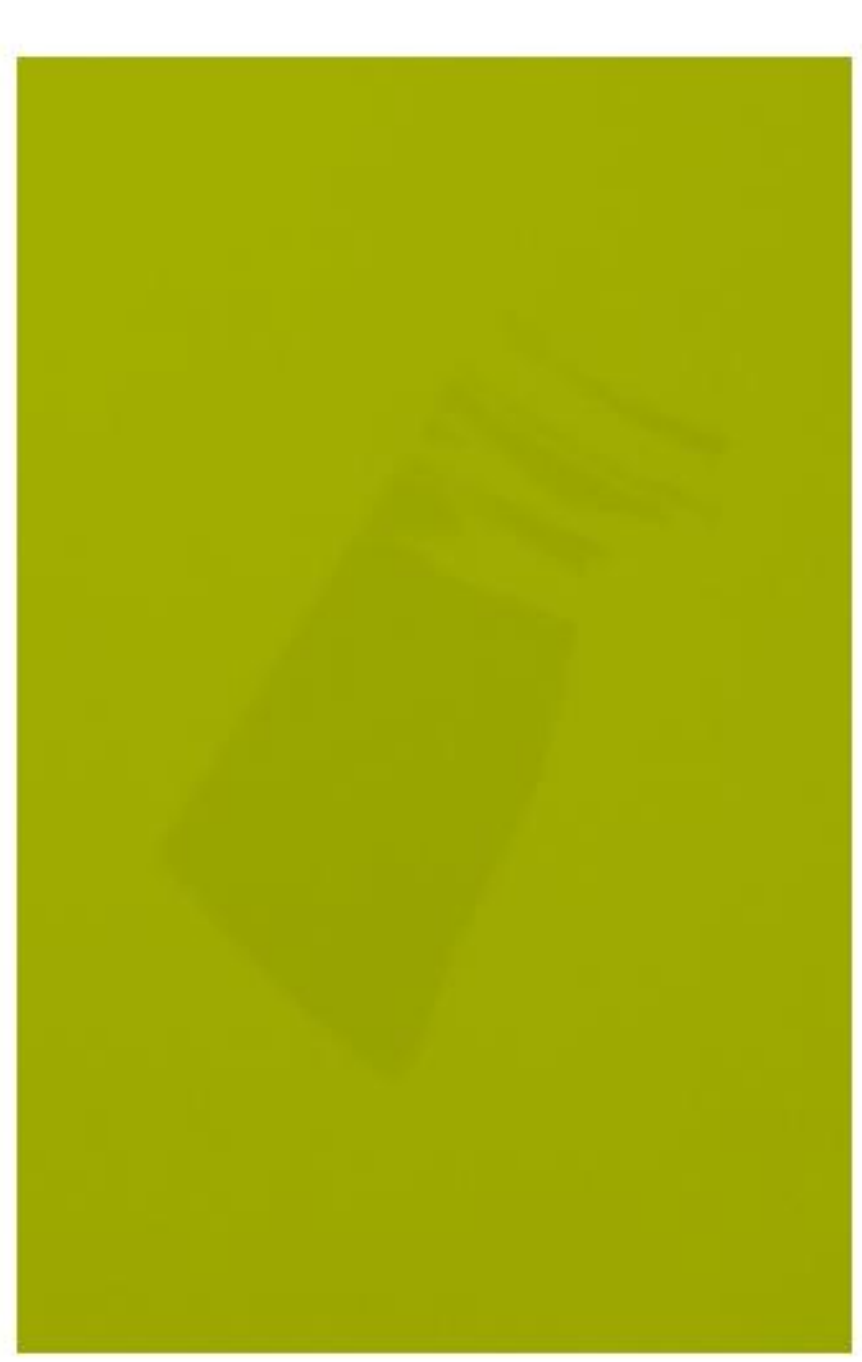

**Supplementary Figure S1) Exfoliated ultra-thin a-$RuCl_3$ flakes on a)PPC and b)$Si/SiO_2$ wafer.**

**Alt text: Optical images of exfoliated ultrathin a-RuCl3 flakes on PPC (left image) and on a Si/SiO2 substrate (right image).**

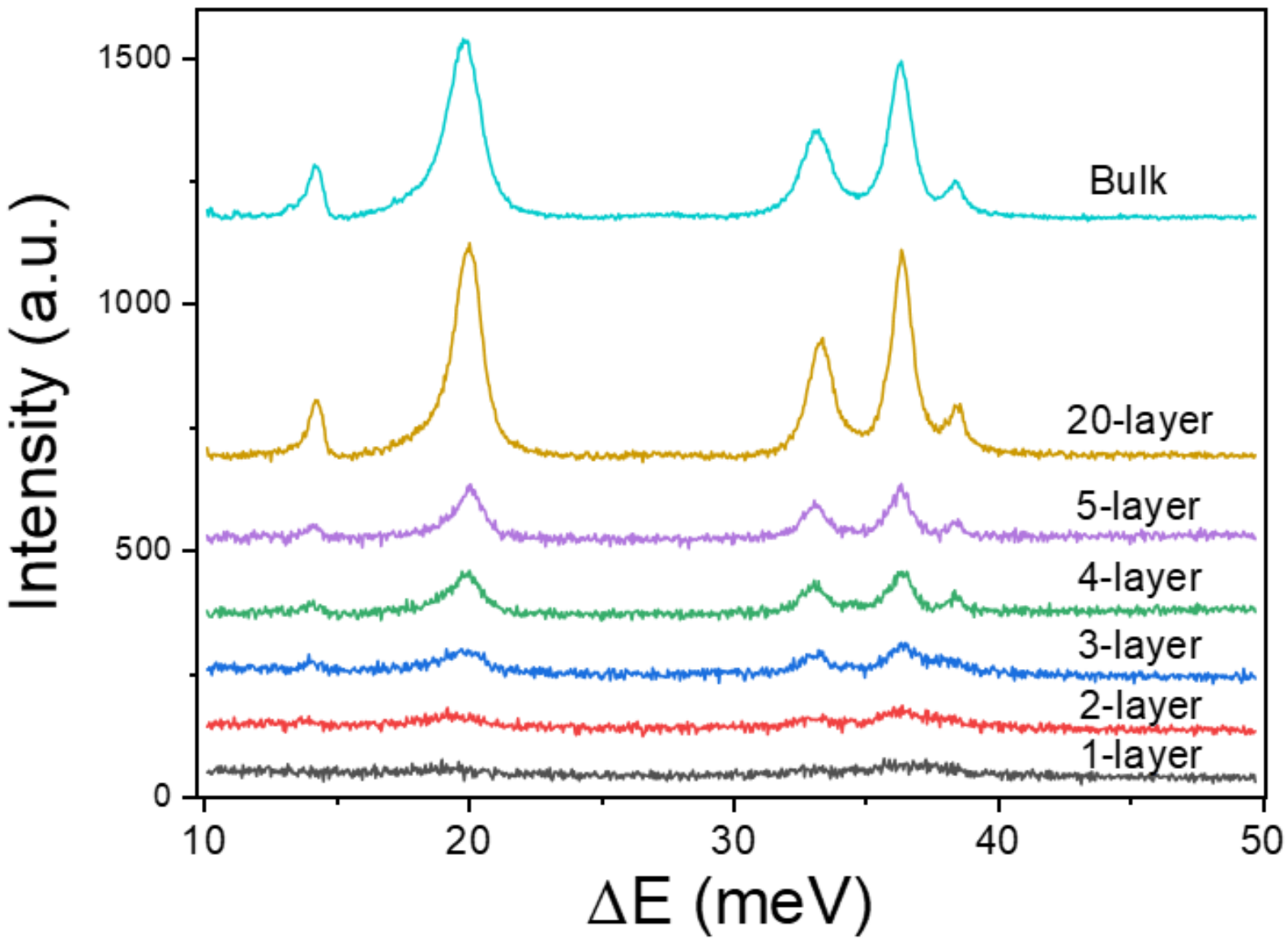

**Supplementary Figure S2) Layer-dependent Raman spectra of a-$RuCl_3$, bulk and exfoliated flakes down to a mono-layer thickness, with previously reported vibrational modes present. We particularly note the lower-energy vibrational modes at 14 meV and 20 meV energy, which we can resolve in our tunnelling study.**

**Alt text: Layer-dependent Raman spectroscopy of a-$RuCl_3$ films, showing 5 vibrational modes up to an energy value of 50 meV.**

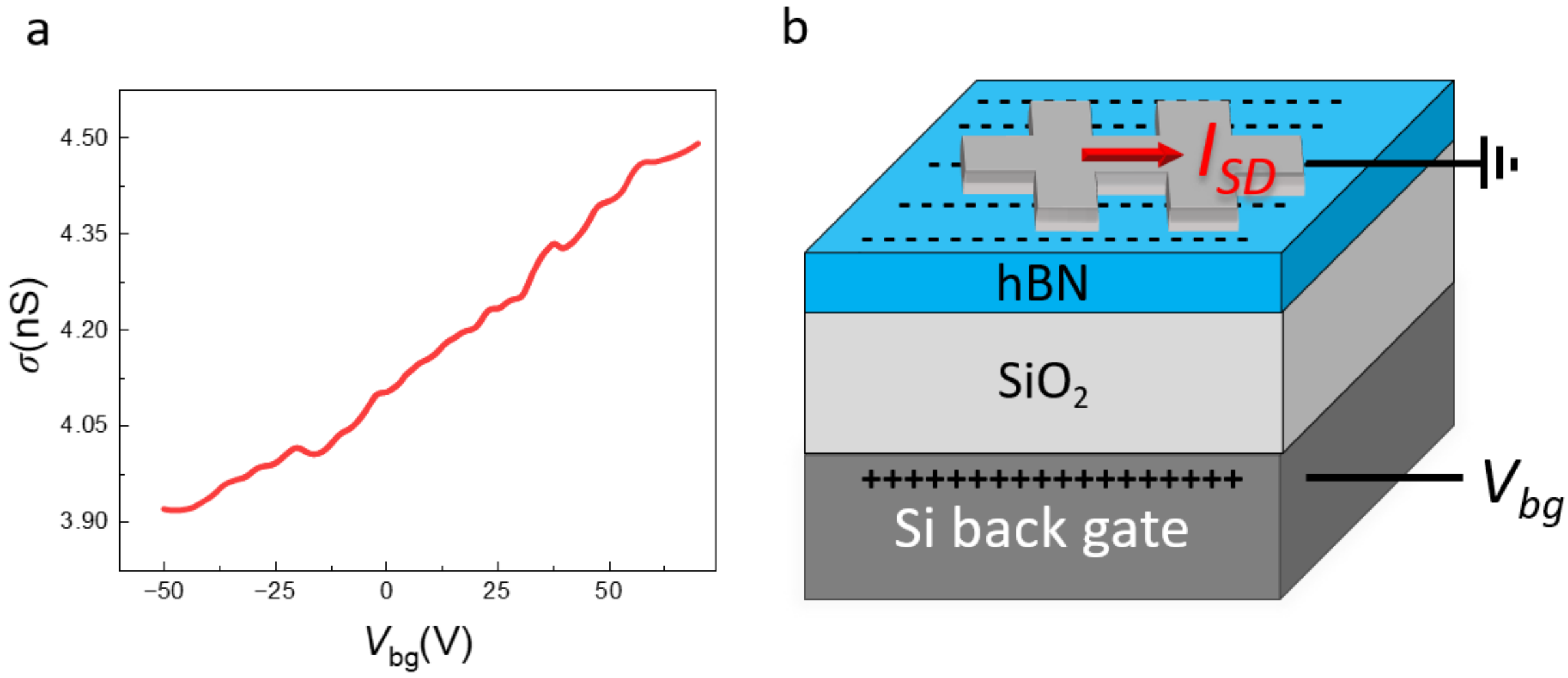

**Supplementary Figure S3) a) Gate dependence of conductivity on a tri-layer a-$RuCl_3$ device showing an n-type electric field effect. b) Schematic of the electrostatic gating set-up, where a gate bias is applied between the Si substrate and the a-$RuCl_3$ film, across the $SiO_2$ and bottom hBN dielectrics.**

**Alt text: Backgate voltage dependence of 4-probe conductance measured at room temperature (left figure) and schematic of the heterostructure and the electrostatic gate effect (right figure).**

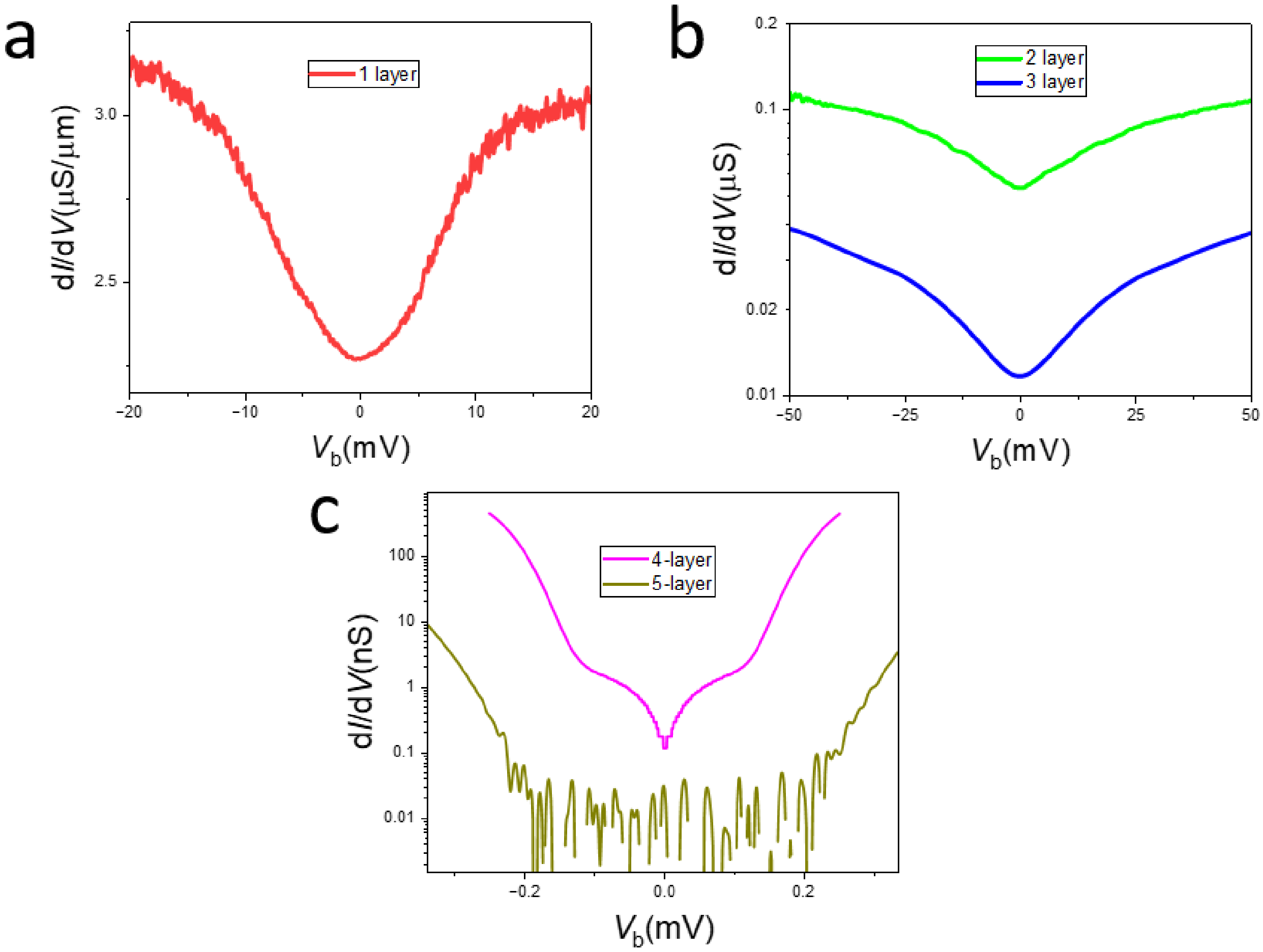

**Supplementary Figure S4) Differential conductance curves of various tunnel junction devices up to a 5-layer thickness of a-$RuCl_3$ with screening onset beyond 4-layer thickness films.**

**Alt text: Layer-dependent difference conductance as a function of DC bias for mono- to five-layer tunnelling devices of a-RuCl3.**

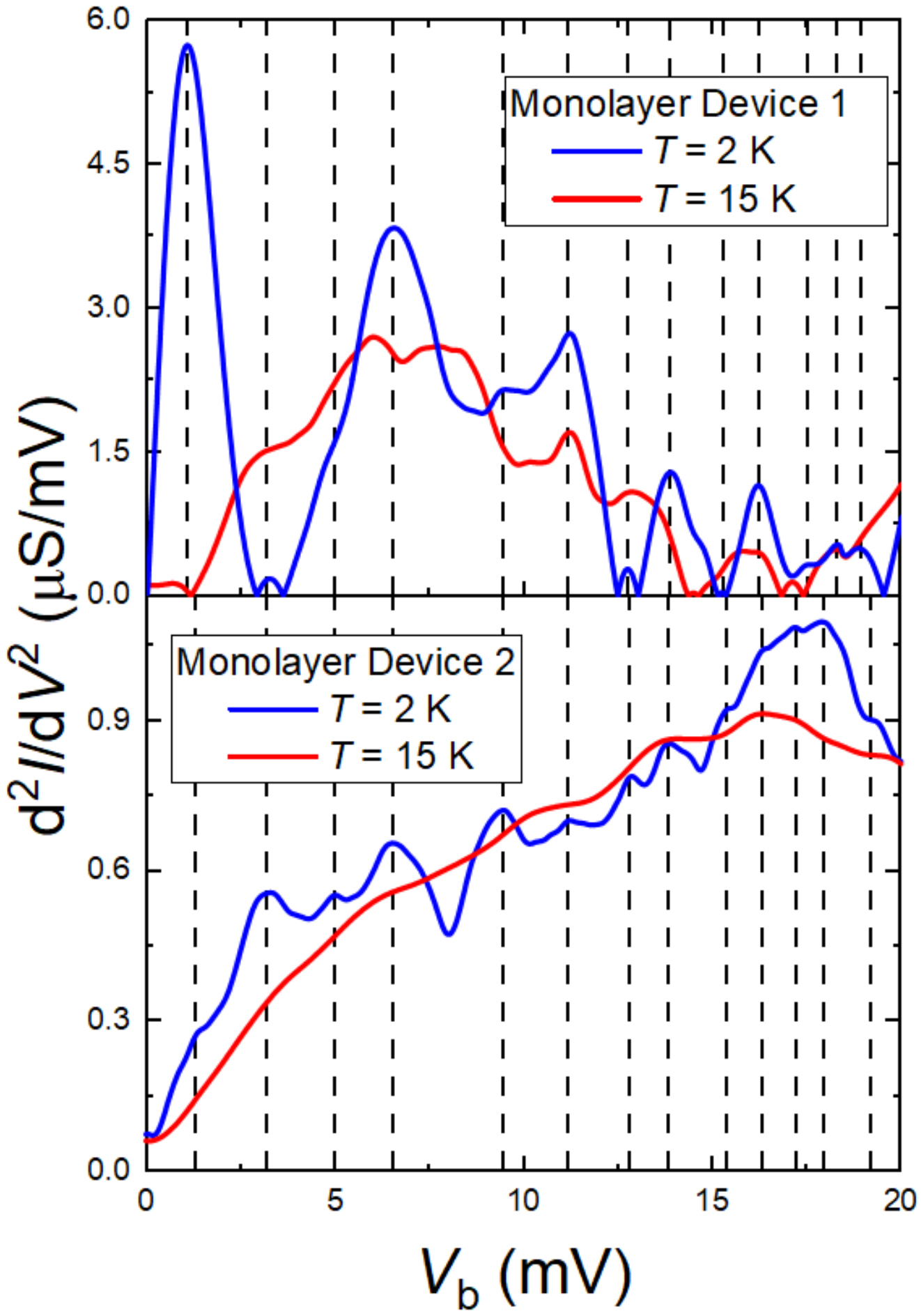

**Supplementary Figure S5) Derivative of differential conductance measured on the first monolayer tunnelling device (top) and second monolayer tunnelling device (bottom) below the Neel temperature (blue) and above the Neel temperature (red). The single modes resolved within the 20 mV DC bias range are highly reproducible, as shown with dashed lines, despite differing appearance of the continuum.**

**Alt text: Comparison of the bias dependence of the derivative of the differential conductance in two-monolayer a-RuCl3 tunnelling devices.**

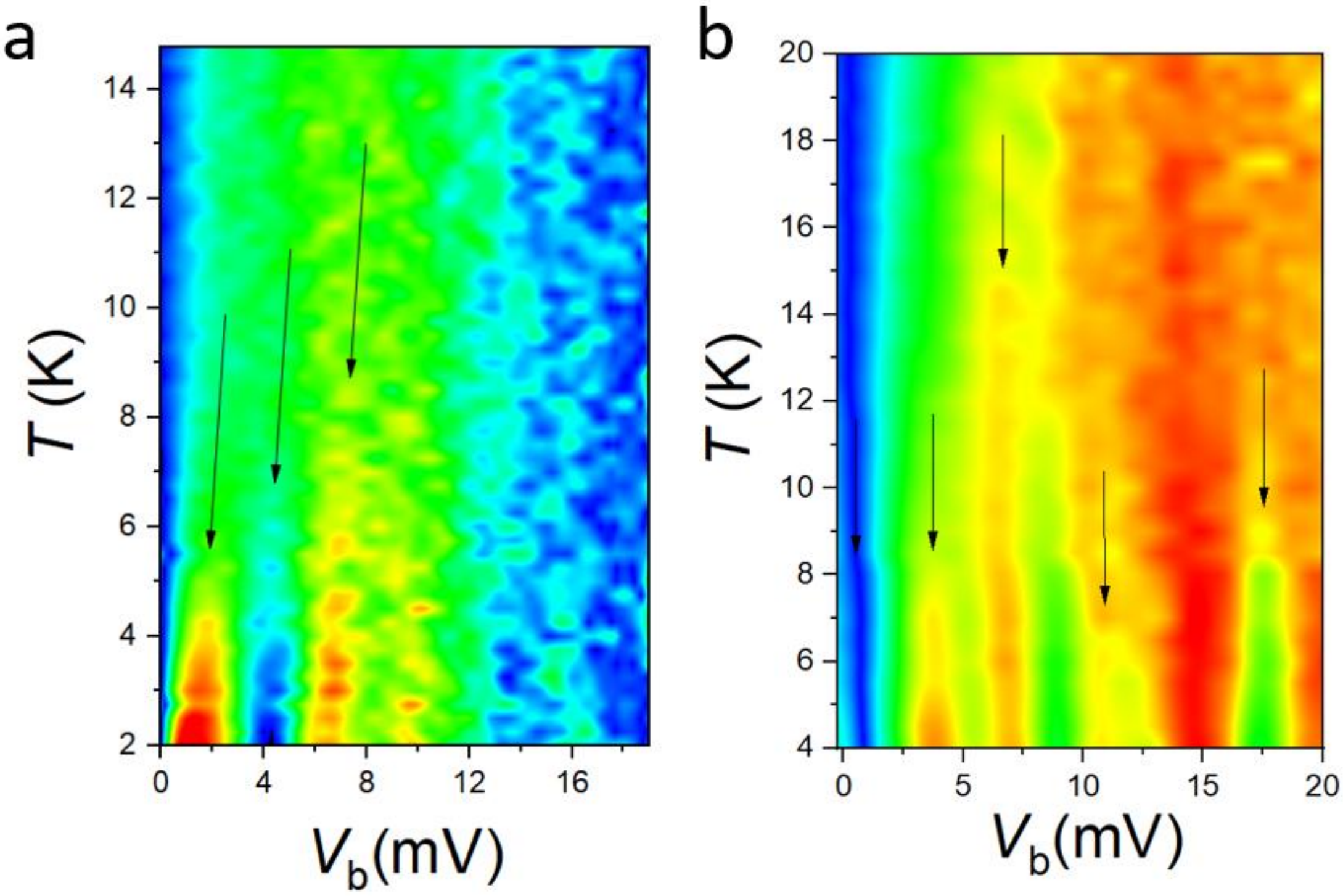

**Supplementary Figure S6) Temperature dependence map of derivative of differential conductance, a) – monolayer device 1, b)-monolayer device 2, showing disappearance of single magnon mode assisted tunnelling features above magnetic transition temperature and merging of phonon related features around higher bias.**

**Alt text: Comparison of temperature dependence of derivative of differential conductance maps as a function of DC bias in two monolayer devices, showing magnon and phonon-assisted features.**

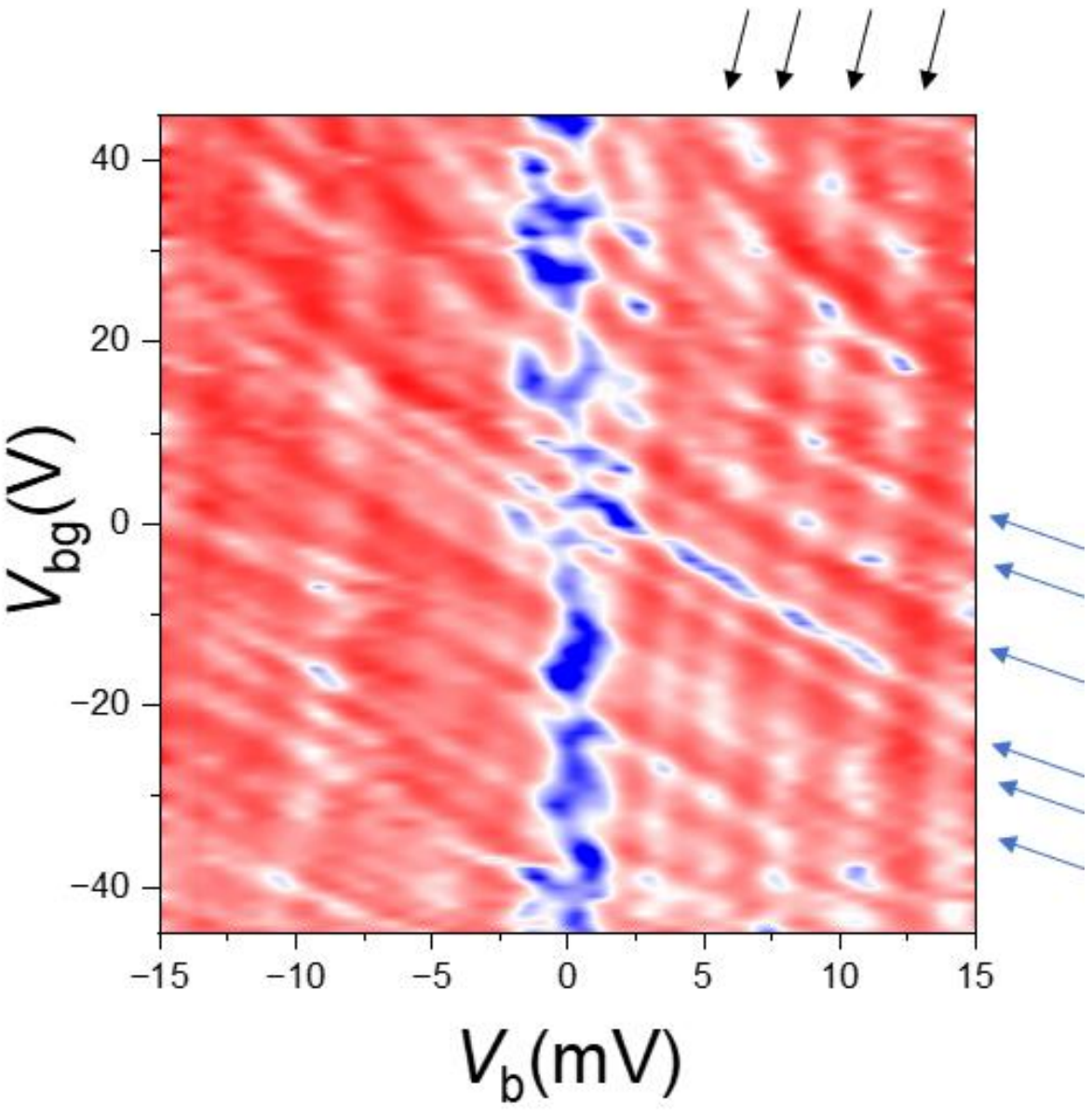

**Supplementary Figure S7) Backgate dependence map of second derivative of tunnel current as a function of DC bias on a bilayer device showing horizontal features (pointed by black arrows) and diagonal features (pointed by blue arrows).**

**Alt text: Backgate dependence map of derivative of differential conductance.**

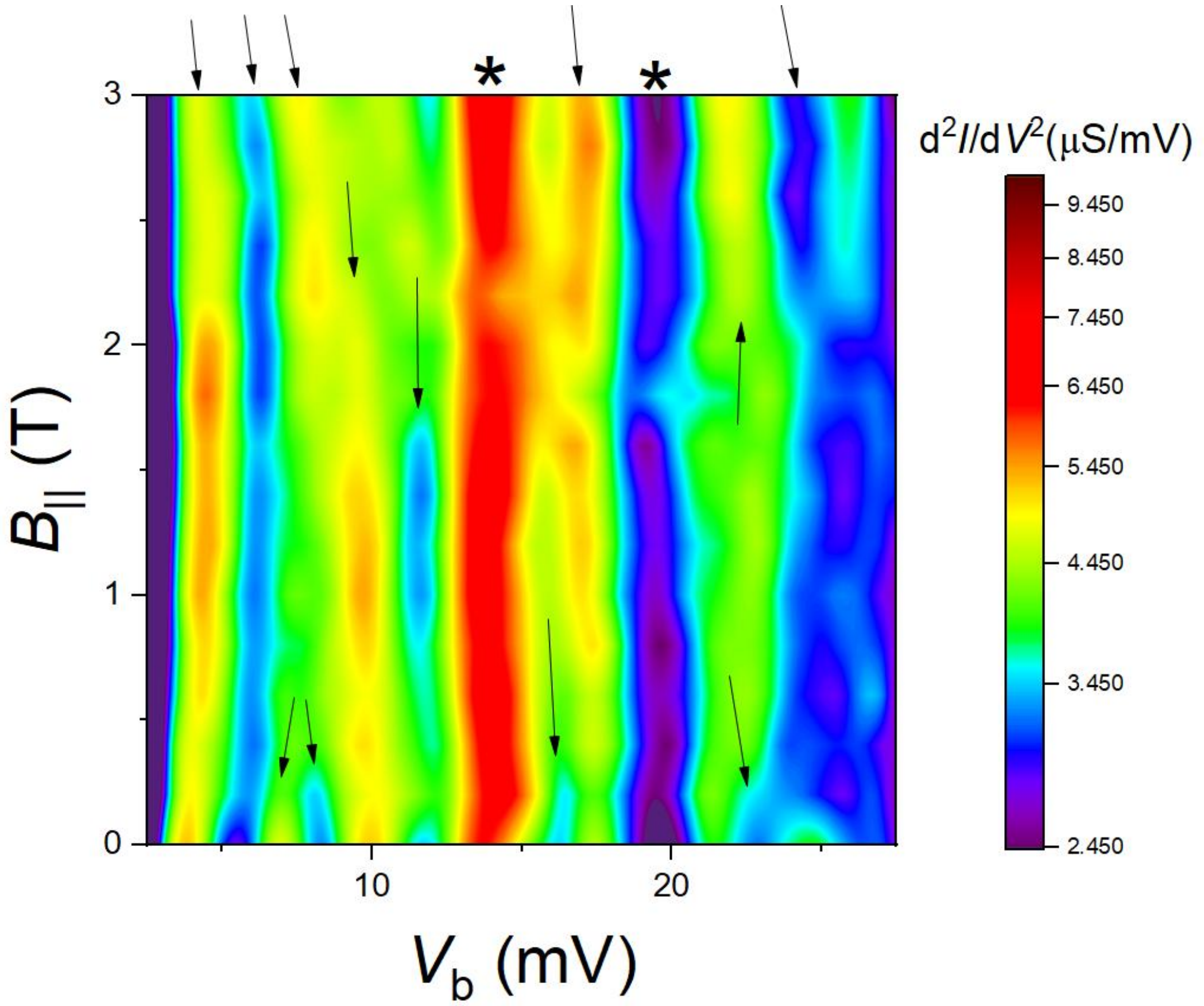

**Supplementary Figure S8) In-plane magnetic field dependence of derivative of differential conductance measured at a temperature of 2 K on a bilayer tunnel device, with features corresponding to single magnons influenced by off-diagonal spin Hamiltonian terms highlighted with arrows and features corresponding to phonons highlighted with stars.**

**Alt text: In-plane magnetic field dependence of derivative of differential current as a function of DC bias, enabling magnon modes and phonon modes to be differentiated.**

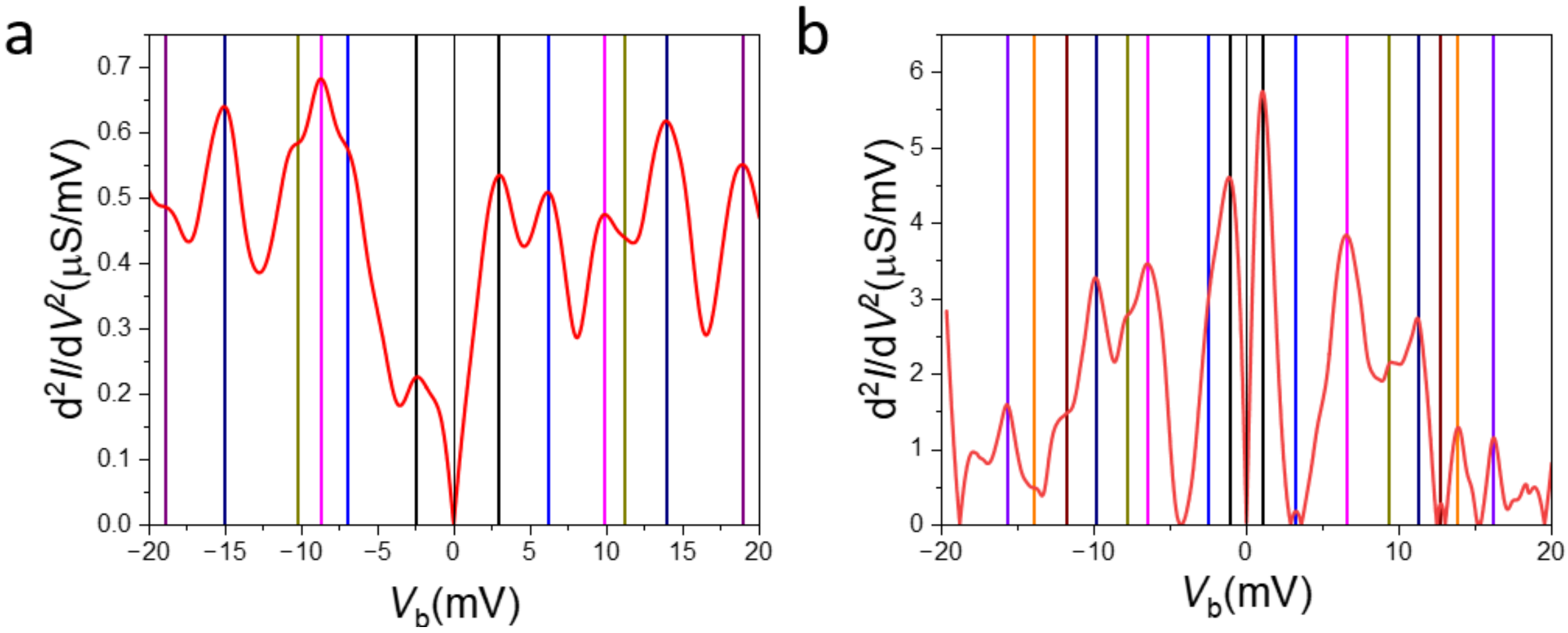

**Supplementary Figure S9) Both positive and negative DC bias plotted the derivative of differential conductance as a function of bias on a) a bilayer tunnelling device and b) a monolayer tunnelling device, showing the same features to be present for opposing DC bias polarities, with peaks corresponding to each other in opposite polarities highlighted with vertical lines of the same colour.**

**Alt text: Comparison of positive and negative bias features on a bilayer (left) and monolayer (right) a-RuCl3 tunnel barrier.**

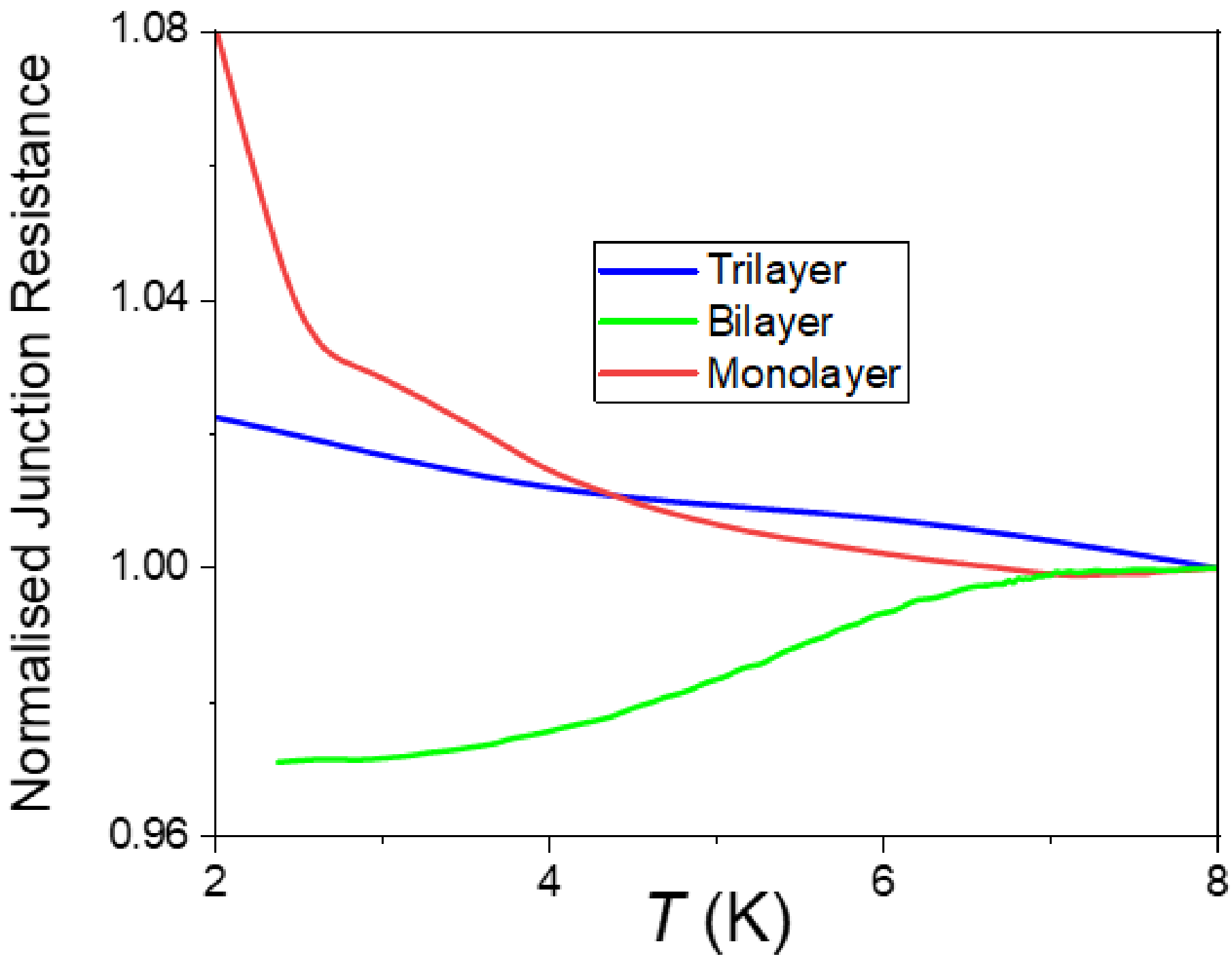

**Supplementary Figure S10) Temperature dependence of normalised tunnel junction resistance of mono-, bi-, and tri-layer a-RuCl3 devices below the Neel temperature.**

**Alt text: Normalised junction resistance as a function of temperature on monolayer, bilayer and trilayer tunnel barriers.**